**Title Page**

**Title**: Beyond Dose in Boron Neutron Capture Therapy: Cellular $^{10}B$-Capture Statistics and Microdosimetric Context in Effect Prediction

**Authors**: Shuichi Furuya[1] and Atsushi Fujimura[2]

[1] Particle Beam Therapy Research Institute, Inc., c/o Tanaka Law Office, Kanda Nishiki-cho 1-21-1, Chiyoda-ku, Tokyo 101-0054, Japan.

[2] Department of Molecular Physiology, Kagawa University Faculty of Medicine, 1750-1 Ikenobe, Miki-cho, Kita-gun, Kagawa 761-0793, Japan.

**Corresponding author:**

Atsushi Fujimura, M.D., Ph.D.

Department of Molecular Physiology, Kagawa University Faculty of Medicine, Graduate School of Medicine

1750-1 Ikenobe, Miki-cho, Kita-gun, Kagawa 761-0793, Japan.

Tel: +81-87-891-2095

e-mail: fujimura.atsushi@kagawa-u.ac.jp

**Short title:** Cellular $^{10}B$-Capture Statistics in BNCT

**Structured Abstract**

**Background:** Boron neutron capture therapy (BNCT) produces high-linear-energy-transfer, short-range particles through the $^{10}B(n,\alpha)^{7}Li$ reaction. Conventional descriptors, including absorbed dose, compound biological effectiveness (CBE)-weighted dose, and photon-isoeffective dose ($D_{isoE}$), remain clinically useful but compress cellular heterogeneity, compartmental boron localization, and stochastic reaction occurrence into macroscopic summaries.

**Purpose:** To propose a hierarchical framework that places cell-associated $^{10}B$-capture reaction statistics upstream of microdosimetric effect modeling and links clinical observations to cellular-scale determinants of BNCT response.

**Methods:** We performed a narrative synthesis of published work on BNCT microdosimetry, CBE and $D_{isoE}$, stochastic microdosimetric kinetic (SMK) modeling, PHITS-based cellular simulations, boron imaging, tissue microstructure, and boron-agent characterization. These elements were reorganized into observational, latent-cellular, reaction-count, and biological-effect layers.

**Results:** The framework separates capture-reaction frequency from microdosimetric quality. Imaging, pathology, agent-specific data, and neutron-field information constrain latent variables such as cellular boron burden, compartmental localization, and cellular geometry. These variables determine expected reaction burdens and distributions of realized reaction counts, which are then translated into cell survival, tumor control, or normal-tissue injury by microdosimetric models. Published cellular-scale simulations support this separation; at a fixed macroscopic boron concentration of 60 ppm, predicted $D_{isoE}$ varied from 7.4 to 32.7 Gy according to cellular morphology and boron distribution.

**Conclusions:** Cell-associated reaction-count statistics provide an interpretable inferential bridge between macroscopic measurements and established BNCT effect models. Their integration with microdosimetry may support spatially resolved, uncertainty-aware prediction of tumor response and normal-tissue risk without replacing existing CBE-, $D_{isoE}$-, SMK-, or PHITS-based approaches.

**Keywords**: boron neutron capture therapy; microdosimetry; $^{10}$B-capture reaction count; hierarchical modeling; treatment effect prediction; imaging; tumor heterogeneity; precision oncology

**Highlights:**

- Averaged dose metrics miss key cellular determinants of BNCT response.
- Cell-associated capture-reaction statistics form an upstream stochastic layer.
- Microdosimetry defines the quality of capture-induced energy deposition.
- Imaging and pathology constrain latent cellular variables in BNCT.
- Hierarchical inference enables spatially explicit BNCT effect modeling.

## Main Text

### 1. Why a Simple Dose–Effect Relationship Does Not Hold in BNCT

#### 1.1. Why the Gy Concept Has Worked in Conventional Radiotherapy, and Why BNCT Requires a Different Level of Description

In external beam radiotherapy, absorbed dose expressed in Gy has long served as the principal quantity for describing biological effect [1,2]. Although radiobiological responses are modified by radiation quality, dose rate, fractionation, oxygenation, DNA repair, and cell-cycle status, Gy remains a practical macroscopic descriptor because photon-induced energy deposition can, to a first approximation, be averaged across tissue volumes [1,2].

BNCT presents a different conceptual problem. Its therapeutic rationale is intrinsically cellular: cytotoxicity arises only where $^{10}B$ atoms encounter neutrons and generate short-range charged particles. Nevertheless, BNCT effect prediction still relies heavily on tissue-averaged quantities such as absorbed dose and CBE-weighted dose. This creates a mismatch between the cellular scale at which the decisive reactions occur and the macroscopic scale at which treatment effect is commonly summarized.

Modern oncology increasingly recognizes that treatment response depends on intratumoral heterogeneity, cellular state, tissue microenvironment, and spatial organization. These considerations are especially relevant to BNCT, because variation in boron delivery and tissue architecture directly determines where capture reactions can occur. Faithful BNCT effect prediction therefore requires a level of description that preserves cellular and spatial heterogeneity rather than relying exclusively on bulk-tissue averages [3–5].

### 1.2. BNCT Is Defined by Localized High-LET, Short-Range Capture Reactions

The principal reaction in BNCT is $^{10}B(n,\alpha)^{7}Li$, triggered when $^{10}B$ captures a thermal neutron [4,5]. The resulting α particle and $^{7}Li$ nucleus have high linear energy transfer and ranges of only a few micrometers. Consequently, biological effect depends not only on how much energy is deposited, but also on where that deposition occurs relative to radiosensitive cellular structures [4–7].

This localization dependence has long been recognized. Ono emphasized that the biological effectiveness of BNCT is closely related to the microscopic arrangement of capture reactions [3]. Early Monte Carlo work by Gabel and colleagues showed that biological outcome depends on the microscopic distribution of boron and cannot be adequately predicted from average dose alone [6]. Santa Cruz and Zamenhof likewise established microdosimetry as a foundational component of BNCT radiobiology rather than an ancillary correction to boron dose [7].

A capture reaction occurring in or near the nucleus is therefore not equivalent to one occurring in the cytoplasm or extracellular space, even when both contribute to the same macroscopic absorbed dose. Averaged quantities remain useful for treatment description, but they do not retain the spatial context of the underlying reactions. This localization dependence is a primary reason why conventional dose–effect concepts cannot be transferred to BNCT without additional cellular-scale information [3–7].

### 1.3. Intercellular Heterogeneity, Compartmental Localization, and the Pitfall of Bulk Tissue Analysis

Boron distribution within tumors is heterogeneous. Cellular uptake may vary because of differences in perfusion, drug accessibility, transporter expression, cellular state, necrosis, and local tissue organization. Ono and colleagues demonstrated that heterogeneous boron microdistribution influences tumor cure, while subsequent modeling by Sato and colleagues showed that both intracellular and intercellular heterogeneity materially affect BNCT effectiveness [8,9].

The relevant variation is not limited to the amount of boron associated with each cell. Its compartmental localization and the geometry of the target cell also matter. Ono and colleagues showed that nucleocytoplasmic ratio and cell size can predict the biological effectiveness of the $^{10}B(n,\alpha)^{7}Li$ component [10]. Takeno and colleagues further identified nuclear diameter, nuclear-to-cytoplasmic ratio, and heterogeneous boron distribution as determinants of capture-reaction sensitivity [11]. Thus, cellular geometry, tissue architecture, and nuclear, cytoplasmic, pericellular, or extracellular boron localization influence how capture reactions are translated into biological injury.

These limitations also apply to the preclinical development of boron-delivery agents. Bulk tumor boron concentration is an indispensable first-line measurement, but it is necessary rather than sufficient evidence of therapeutic efficacy. A high average concentration in a tumor homogenate does not establish delivery to malignant cells, uniformity across viable tumor regions, or localization near radiosensitive targets. Preclinical evaluation should therefore complement bulk boron concentration with measurements or estimates of cellular uptake heterogeneity, compartmental localization, and spatial correspondence between boron distribution and viable tumor-cell populations.

### 1.4. What CBE, RBE, and $D_{\mathrm{isoE}}$ Have Achieved, and Where They Remain Incomplete

RBE and CBE were introduced to account for radiobiological features that are not represented by absorbed dose alone. CBE has been particularly valuable as a practical measure that incorporates, in aggregate, differences among boron compounds, tissues, and biological endpoints [3]. However, CBE is not a universal fixed coefficient. Its value reflects multiple microscopic and tissue-dependent factors, including boron distribution and structural context. Its practical strength therefore derives from compressing these determinants into a tractable summary, but that compression also defines its limitation.

González and Santa Cruz subsequently introduced photon-isoeffective dose, $D_{\text{isoE}}$, to express BNCT exposure as the photon dose expected to produce an equivalent biological effect [12]. Sato and colleagues further advanced BNCT effect modeling through stochastic microdosimetric kinetic approaches that incorporate nucleus- and domain-level specific-energy distributions, heterogeneity in $^{10}$B distribution, radiation-component interactions, dose dependence, and saturation or overkill [9,13,14].
These developments represent substantial progress beyond fixed weighting-factor approaches. Nevertheless, their practical outputs are often expressed as dose-equivalence or effect-equivalence summaries. The present article seeks to make explicit the upstream reaction-count layer from which such downstream effects arise. We focus on the expected $^{10}$B-capture reaction burden associated with target cells and on the spatial and compartmental context in which those reactions occur.

This proposal is therefore not intended to reject CBE-, $D_{\text{isoE}}$-, or SMK-based approaches. Rather, it reorganizes them within a hierarchy in which cell-associated reaction-count statistics provide a physically interpretable upstream variable and microdosimetry determines how those reactions are converted into biological effect.

### 1.5. Recent Implementation Studies Have Sharpened the Problem

Recent SMK- and PHITS-based implementation studies have made the importance of microscopic inputs particularly clear. Shigehira and colleagues developed LISMEC, which uses precomputed cellular-scale PHITS simulations to interpolate SMK parameters across different cellular morphologies and boron-distribution conditions [15]. Even when the macroscopic $^{10}$B concentration was fixed at 60 ppm, predicted $D_{\mathrm{isoE}}$ ranged from 7.4 to 32.7 Gy depending on boron uptake ratio, cell occupancy, nuclear-to-cell area ratio, and nuclear size [15].

This result demonstrates that explicit representation of cellular morphology and boron distribution can profoundly alter the downstream effect metric despite an identical macroscopic concentration. It therefore provides direct support for treating cell-associated reaction burden and its spatial context as upstream determinants of BNCT effect. The relevant uncertainty lies not only in neutron fluence or mean boron concentration, but also in the cellular architecture and microdistribution that determine where reactions occur and how their energy is deposited.

### 1.6. Problem Setting of the Present Article

Taken together, averaged dose descriptors remain clinically useful but do not explicitly represent the discrete, heterogeneous, and compartment-dependent nature of $^{10}$B-capture reactions. BNCT effect prediction should therefore place cell-associated reaction-count statistics upstream of microdosimetric effect modeling rather than begin with mean dose and subsequently add corrections.

Building on prior work in Monte Carlo simulation, microdosimetry, photon-isoeffective dose, heterogeneity-aware modeling, and cellular-scale implementation [6,7,9,12–15], we propose a hierarchical framework in which observable clinical and experimental data constrain latent cellular variables, these variables determine reaction-count distributions, and microdosimetry assigns biological consequence. The following sections develop this capture-reaction-frequency–microdosimetric-quality architecture in detail.

## 2. Mean Reaction-Count Models and Microdosimetry

### 2.1. $^{10}$B-Capture Reaction Count as an Upstream Quantity in BNCT

The principal terms and notation used throughout this article are summarized in Box 1. The most direct physical quantity underlying BNCT is the number of $^{10}B(n,\alpha)^{7}Li$ reactions associated with each target cell. This quantity should not be restricted to reactions occurring within the cell nucleus or even within the intracellular space. Because the emitted α particle and $^{7}Li$ nucleus travel several micrometers, reactions in the cytoplasm, membrane-proximal region, pericellular space, or nearby extracellular compartment may contribute to energy deposition in radiosensitive cellular domains [6,7,9,15–17].

A compartment-resolved notation is therefore appropriate. For target cell $i$ and spatial compartment $c$, the expected $^{10}$B-capture reaction burden can be written in simplified form as:

$$\lambda_{i,c} = \Phi_{i,c} \cdot \sigma_B \cdot n_{B,i,c},$$

where $\Phi_{i,c}$ is the time-integrated neutron fluence experienced by the compartment, $\sigma_B$ is the energy-averaged cross section for the $^{10}B(n,\alpha)^{7}Li$ reaction, and $n_{B,i,c}$ is the number of $^{10}$B atoms associated with that compartment. The total target-cell-associated reaction burden is then:

$$\lambda_i = \sum_c \lambda_{i,c}.$$

When the neutron energy spectrum must be represented explicitly, the corresponding expression is:

$$\lambda_{i,c} = n_{B,i,c} \cdot \int \Phi_{i,c}(E) \cdot \sigma_B(E)\, dE$$

The simplified formulation is sufficient for the present conceptual framework, provided that the spatial meaning of $\lambda_{i,c}$ is stated explicitly [7,16]. Different compartments may contribute unequally to biological effect because their reactions differ in proximity to the nucleus and other radiosensitive structures.

Reaction burden is therefore an upstream capture-reaction-frequency variable, not a replacement for absorbed dose or microdosimetry. Dose and equivalent-effect metrics summarize the aggregate consequences of reaction frequency, localization, and microscopic energy deposition. Explicitly representing reaction burden preserves a physical structure that those downstream summaries necessarily compress [6,7,9,12,14,15].

### 2.2. BNCT Is Inherently a Stochastic Process of Discrete Capture Reactions

The expected reaction burden, $\lambda_{i,c}$ or $\lambda_i$, is not identical to the number of reactions realized in an individual target cell. Conditional on a specified cellular state—including boron burden, compartmental localization, tissue geometry, and neutron fluence—the realized reaction count may be approximated as a Poisson-distributed variable:

$$P\left(k_{i,c} = k \middle| \lambda_{i,c}\right) = \frac{\lambda_{i,c}^k e^{-\lambda_{i,c}}}{k!}.$$

After compartmental information is collapsed into a target-cell-associated reaction burden:

$$P(k_i = k|\lambda_i) = \frac{\lambda_i^k e^{-\lambda_i}}{k!}$$

Thus, an average of one reaction per cell does not imply that every cell undergoes exactly one reaction. Some cells may undergo none, while others undergo one or several. The expected burden describes the mean of a distribution of discrete reaction histories.

At the population level, $\lambda_i$ itself is heterogeneous. It varies among cells according to boron uptake, compartmental localization, cellular and nuclear geometry, cell packing, tissue architecture, and local neutron fluence. The reaction-count distribution across a tumor should therefore not be represented by a single Poisson distribution with one uniform value of $\lambda$. It is more appropriately described as a mixture of conditional distributions:

$$p(k) = \int P(k|\lambda)p(\lambda)\,d\lambda,$$

where $p(\lambda)$ represents the latent distribution of expected reaction burdens generated by biological, anatomical, and physical heterogeneity.

BNCT stochasticity consequently arises at two levels. The first is the discreteness of capture reactions: even for a fixed $\lambda_i$, the realized count fluctuates. The second is heterogeneity in the expected burden itself: different cells have different values of $\lambda_i$. Averaged dose metrics tend to conceal both sources of variability.

This distinction has direct implications for tumor control. Cells with low $\lambda_i$ may undergo few or no biologically relevant capture reactions even when the average tumor dose appears adequate. If such cells are spatially clustered within viable or clonogenic tumor regions, they may contribute disproportionately to local recurrence. Conversely, treatment conditions with similar average dose may generate different reaction-count distributions and therefore different probabilities of eliminating all clonogenic cells.

This formulation complements existing microdosimetric and SMK-based models by separating stochastic reaction occurrence from the stochasticity of microscopic energy deposition and biological response [7,9,14,16].

### 2.3. Microdosimetry Defines the Biological Consequence of Capture Reactions

Reaction count alone does not determine BNCT effect. The biological consequence of a $^{10}$B-capture reaction depends on its location relative to radiosensitive structures and on whether the emitted α particle or $^{7}$Li nucleus deposits energy within the nucleus or its subdomains. Two reactions associated with the same target cell may therefore differ substantially in biological potency.

Microdosimetry describes this microscopic energy-deposition structure. Santa Cruz and Zamenhof emphasized that boron dose cannot serve as a self-sufficient descriptor once the spatial distribution of energy deposition is considered [7]. In the present framework, microdosimetry does not primarily determine how frequently capture reactions occur. It determines the quality and biological consequence of the energy deposition resulting from those reactions [7,16,18,19].

SMK-based BNCT models provide a concrete implementation of this principle. They translate nucleus- and domain-level specific-energy distributions into surviving fractions while accounting for intra- and intercellular $^{10}$B heterogeneity, interactions among radiation components, dose dependence, saturation, and overkill [9,13,14].

BNCT effect is therefore determined neither by average dose alone nor by reaction count alone. Cell-associated reaction statistics describe capture-reaction frequency, whereas microdosimetry describes how those reactions are converted into biological injury.

## 2.4. BNCT Effect Is Best Understood as Capture-Reaction Frequency Coupled to Microdosimetric Quality

BNCT effect can be organized into two complementary layers. The first is capture-reaction frequency: how many $^{10}B(n,\alpha)^{7}Li$ reactions are associated with a target cell or tissue region. The second is microdosimetric quality: how the charged particles generated by those reactions deposit energy within radiosensitive domains and how that deposition is translated into biological injury.

Reaction-count modeling and microdosimetry are therefore complementary rather than competing. Reaction-count modeling describes the stochastic occurrence of capture reactions, whereas microdosimetry describes their spatial energy-deposition structure and biological consequence. Existing $D_{\mathrm{isoE}}$- and SMK-based approaches can thus be situated downstream of, rather than displaced by, the reaction-count layer [9,12–14].

This relationship may be represented schematically by linking the conditional reaction-count distribution to a survival function. For target cell $i$, the probability of cell death can be written as:

$$P_{death,i} = 1 - \sum_{k=0}^{\infty} P(k_i = k|\lambda_i) \cdot S_{i(k,\theta_i)}$$

where $P(k_i = k \mid \lambda_i)$ is the conditional distribution of realized capture-reaction counts and $S_i(k, \theta_i)$ is the survival probability after $k$ relevant reactions under microdosimetric and biological context $\theta_i$.

The parameter $\theta_i$ may include reaction-site proximity to radiosensitive domains, nuclear and cellular geometry, domain-level specific-energy distributions, repair capacity, cell-cycle state, and overkill. This expression is not intended as a finished survival model.

Rather, it makes explicit the conceptual distinction between how many reactions occur and how biologically consequential those reactions are.

This organization shifts BNCT effect prediction away from assigning increasingly refined weights to averaged dose and toward integrating reaction-count statistics with microdosimetric-quality modeling.

**2.5. Cellular Morphology and Localization Shape Microdosimetric Quality**

Microdosimetric quality is strongly influenced by cellular morphology and boron localization. Ono and colleagues showed that the biological effectiveness of the $^{10}B(n,\alpha)^{7}Li$ component can be predicted from nucleocytoplasmic ratio and cell size [10]. Takeno and colleagues similarly identified nuclear diameter, nuclear-to-cytoplasmic ratio, and heterogeneous boron distribution as determinants of capture-reaction sensitivity [11].

These findings indicate that the lethality of a capture reaction is not fixed by nuclear physics alone. A reaction occurring within or near the nucleus is not equivalent to one occurring in a distant extracellular compartment, even if both contribute to the same macroscopic boron dose. Likewise, cells with the same expected reaction burden may differ in survival because of differences in nuclear size, cytoplasmic volume, cell packing, or compartmental boron localization.

LISMEC further demonstrated the quantitative importance of these parameters. By combining cellular-scale PHITS simulations with SMK-based interpolation, Shigehira and colleagues showed that cell occupancy, nuclear geometry, and intracellular versus extracellular boron distribution can substantially alter predicted $D_{\mathrm{isoE}}$, even at a fixed macroscopic $^{10}B$ concentration [15].

The proposition that the same number of reactions necessarily produces the same biological effect is therefore untenable. Reaction-count statistics determine the distribution of capture-reaction burden, whereas cellular morphology and localization determine how that burden is translated into microscopic energy deposition and biological injury [9–11,15].

### 2.6. Position of the Present Framework Relative to $D_{\mathrm{isoE}}$ and SMK

The present framework is a reorganization, not a replacement, of existing BNCT effect models. $D_{\mathrm{isoE}}$ provides a theoretically grounded equivalent-effect quantity, while SMK-based approaches translate microscopic specific-energy distributions into biologically interpretable survival estimates [9,12–14]. PHITS supplies detailed particle-transport and energy-deposition calculations, and LISMEC makes cellular-scale SMK parameter estimation more practical across different morphologies and boron-distribution scenarios [15].

The novelty of the present proposal therefore lies neither in introducing stochasticity nor in introducing microdosimetry into BNCT. Both are already integral to established models. Rather, the contribution is to make cell-associated reaction-count statistics explicit as an upstream latent layer connecting clinical observations to downstream microdosimetric effect modeling.

Imaging, pathology, boron-agent characterization, and neutron-field information constrain latent variables such as cellular boron burden, compartmental localization, tissue architecture, and cellular geometry. These variables determine $\lambda_i$ or $\lambda_{i,c}$, from which reaction-count distributions can be inferred. Those distributions then provide the capture-reaction-frequency input for SMK-, PHITS-, or $D_{\mathrm{isoE}}$-related effect modeling.

The proposed framework is therefore best understood as an inferential bridge. It connects observable macroscopic information to latent cellular reaction statistics and established microdosimetric models, while clarifying the microscopic reaction structure summarized by dose and equivalent-effect metrics. The conceptual relationship between existing BNCT effect models and the present framework is summarized in Table 1.

## 3. Imaging as an Observational Layer, and Its Limits

### 3.1. Imaging Is Indispensable, but It Is Not the Final Descriptor

Imaging is essential for patient selection, lesion assessment, treatment planning, and estimation of boron delivery to tumor and normal tissues [20,21]. In BPA-based BNCT, $^{18}$F-BPA PET provides clinically relevant information on tracer uptake, tumor-to-normal ratios, and lesion-level heterogeneity.

However, imaging does not directly reveal the boron burden of individual malignant cells, cell-to-cell variability, or nuclear, cytoplasmic, pericellular, and extracellular localization. Imaging should therefore be regarded as an observational layer that constrains latent cellular variables rather than as a direct descriptor of biological effect.

### 3.2. PET Reveals Spatial Heterogeneity but Not the Cellular Distribution Within Each Voxel

An important contribution of PET is that it reveals spatial heterogeneity in boron-related uptake at a clinically accessible scale. This is particularly relevant because BNCT treatment planning has often relied on regional or voxel-averaged quantities that may obscure low-uptake areas within a tumor.

Teng and colleagues incorporated heterogeneous boron distribution into BNCT dose calculation using PET-based correction and showed that spatial variation in uptake can alter dose–volume relationships and expose regions masked by mean-value analysis [22]. PET can therefore constrain voxel-level boron heterogeneity more effectively than a single tumor-to-normal ratio or lesion-average value.

Nevertheless, voxel-level correction does not resolve the cellular-scale problem. A voxel with apparently adequate uptake may contain malignant cells with low boron burden, stromal or extracellular accumulation, necrotic regions, or heterogeneous subcellular localization. Conversely, modest voxel-average uptake may coexist with favorable boron localization in a subset of tumor cells.

PET therefore provides an essential spatial constraint, but it does not directly determine the latent distribution of cellular boron burden within each voxel. Its proper role in the present framework is to constrain the range of cellular states compatible with the observed uptake pattern [20–22].

### 3.3. MRI Provides Structural Constraints Rather Than Direct Boron Localization

Conventional clinical MRI does not directly image boron atoms or boron-bearing drugs with sufficient resolution to determine cell-associated $^{10}B$ burden. It can, however, provide information relevant to tissue structure and the cellular microenvironment. Diffusion-based methods, including temporal diffusion spectroscopy, have been explored for estimating features such as cell size, restriction, and compartmental organization [23]. Conductivity-based MRI approaches may also provide estimates related to extracellular volume fraction [24].

These parameters are relevant to BNCT because cellular size, extracellular fraction, tissue compactness, and nuclear density influence the spatial relationship between capture reactions and radiosensitive structures. MRI-derived measurements may therefore help individualize structural priors used to infer the latent cellular state.

The limitation must remain explicit. MRI-derived cell-size or extracellular-volume estimates are not direct measurements of nuclear, cytoplasmic, membrane-proximal, pericellular, or extracellular boron localization. Nor do they determine cell-to-cell variation in boron uptake. MRI should therefore be used as a source of patient- or region-specific structural constraints, not as a direct solution to the microscopic boron-localization problem [23,24].

### 3.4. A Resolution Gap Separates Observable Imaging Data From Effect-Determining Variables

The central imaging limitation in BNCT is a resolution gap. Clinical imaging primarily provides macroscopic or mesoscopic quantities, including voxel-averaged tracer uptake, tumor-to-normal ratio, regional heterogeneity, tissue volume, and structural surrogates. By contrast, the variables that most directly determine BNCT effect are cellular or subcellular: cell-associated boron burden, compartmental localization, nuclear geometry, cell packing, and the spatial relationship between reaction sites and radiosensitive domains.

Cellular-scale simulations make this gap explicit. LISMEC and related SMK-based analyses have shown that cell occupancy, nuclear size, nuclear-to-cell ratio, and intracellular versus extracellular boron distribution can substantially alter predicted $D_{\mathrm{isoE}}$ [15]. Yet these variables are not directly measured by routine PET or MRI.

This does not render imaging inadequate. Rather, it defines its role. PET

constrains boron-related uptake and spatial heterogeneity; MRI constrains tissue structure; pathology constrains cellular and nuclear morphology; and agent-specific studies constrain uptake behavior and microdistribution. Each source is incomplete, but together they restrict the plausible range of latent cellular states [15,19,22–24].

### 3.5. BNCT Imaging Should Be Formulated as an Inverse Problem

Because the variables that determine BNCT effect cannot all be observed directly, imaging should be incorporated into an inverse problem. The objective is not to reconstruct every cell with certainty, but to infer plausible distributions of latent cellular variables from incomplete and spatially averaged observations.

In this formulation, PET constrains voxel-level boron-related uptake and heterogeneity, while MRI provides structural or microenvironmental information. Pathology contributes sampled measurements of cell density, nuclear morphology, stromal content, necrosis, and viable tumor architecture. Boron-agent characterization informs uptake, pharmacokinetics, and compartmental localization.

These complementary observations can be integrated within a hierarchical model to infer cell-associated $^{10}$B burden, compartmental distribution, cellular geometry, and tissue architecture. The inferred latent state then determines expected reaction burdens and reaction-count distributions, which provide the upstream input for microdosimetric and biological-effect modeling.

The limitations of imaging therefore do not preclude quantitative BNCT effect prediction. They establish the need for a multilevel inferential framework in which observable imaging and pathological data constrain, rather than directly define, the cellular variables governing treatment response [9,15,20–24].

## 4. A Hierarchical Statistical Framework for Practical Inference

### 4.1. Why a Hierarchical Model Is Required

The resolution gap described above requires a model that distinguishes observable data from the cellular and subcellular variables that determine BNCT effect. PET, MRI, pathology, boron-agent characterization, and neutron-field calculations provide essential information, but none directly reveals the complete cellular distribution of boron or the spatial context of $^{10}$B-capture reactions.

Existing forward models have shown that cellular morphology and boron microdistribution can substantially alter predicted BNCT effectiveness [9,14,15]. The remaining challenge is therefore an inverse one: to infer these microscopic conditions, with quantified uncertainty, from clinically and experimentally accessible observations.

The proposed framework addresses this problem by separating observational data, latent cellular variables, reaction-count statistics, and biological effect. Figures 1 and 2 summarize this hierarchy and the coupling of capture-reaction frequency to microdosimetric quality.

### 4.2. Four Layers of the Framework

The framework comprises four linked layers.

The first is the observational layer, which includes PET-derived boron-related uptake and heterogeneity, MRI-derived structural surrogates, pathological measures of tissue architecture, boron-agent characteristics, blood boron concentration, and neutron-field information [20–25]. These quantities constrain, but do not directly determine, biological effect.

The second is the latent cellular layer. It contains variables that are biologically important but incompletely observed in clinical practice, including cell-associated boron burden, compartmental localization, cell and nuclear geometry, cell packing, extracellular and stromal fractions, necrosis, and local tissue architecture [9–11,15].

The third is the reaction-count layer. Once the latent cellular state and neutron field are specified, the expected $^{10}$B-capture reaction burden, $\lambda_i$ or $\lambda_{i,c}$, can be estimated for each target cell, compartment, or representative cellular subpopulation. Realized reaction counts are then modeled probabilistically.

The fourth is the biological-effect layer, in which reaction-count distributions are translated into survival, cell death, tumor control, or normal-tissue injury. Microdosimetric models operate at this level by incorporating specific-energy distributions, nuclear and domain-level geometry, saturation, repair, and overkill [9,13,14].

This separation clarifies two questions that are often compressed into a single dose-like quantity: how many capture reactions are associated with a target cell, and how biologically consequential are the resulting microscopic energy depositions?

### 4.3. Priors Are Necessary, but They Must Be Disciplined

Hierarchical inference requires prior distributions for variables that cannot be measured directly. This does not make the framework arbitrary. Reducing microscopic heterogeneity to a single average is itself an assumption, often a stronger one, because it implicitly assumes that cellular and subcellular variation can be neglected.

Priors should therefore be empirically constrained. Cellular and compartmental boron distributions may be informed by uptake assays and spatially resolved measurements such as autoradiography, CR-39 analysis, secondary ion mass spectrometry, or validated surrogate-imaging methods. Histopathology and digital image analysis can constrain cell density, nuclear size, nuclear-to-cell ratio, stromal fraction, necrosis, and viable tumor fraction. Agent-specific studies can inform pharmacokinetics, intracellular retention, and intracellular versus extracellular distribution [25].

The distribution used to represent cellular boron burden within a voxel, whether log-normal, gamma, mixture-based, or otherwise, should be treated as a model-selection question rather than as a fixed biological truth. Candidate distributions should be evaluated according to their consistency with measured voxel averages, tissue data, and experimental microdistribution.

The purpose of the hierarchical framework is not to claim direct observation of the unobservable. It is to represent latent variables explicitly, constrain them empirically, and propagate their uncertainty into reaction-count and biological-effect predictions.

**4.4. A Minimal Mathematical Formulation**

A minimal framework can be written in terms of an observation model, a latent cellular model, a reaction-count model, and a biological-effect model.

Let $y_v$ denote the observations available for voxel or region $v$, including PET-derived uptake, MRI-derived structural information, and pathological measurements when available. Let $Z_v$ denote the latent cellular state, including boron distribution, cellular geometry, and tissue architecture. The observational model may be written schematically as:

$$y_v \sim p(y_v | Z_v, \eta_v),$$

where $\eta_v$ represents measurement noise, imaging resolution, partial-volume effects, and other observation-related parameters.

The latent cellular state determines the expected reaction burden for target cell $i$ and compartment $c$:

$$\lambda_{i,c} = \Phi_{i,c} \cdot \sigma_B \cdot n_{B,i,c},$$

or, when neutron energy dependence is represented explicitly:

$$\lambda_{i,c} = n_{B,i,c} \cdot \int \Phi_{i,c}(E) \cdot \sigma_B(E)\, dE.$$

The realized reaction count is then conditionally stochastic:

$$k_{i,c} \mid \lambda_{i,c} \sim Poisson(\lambda_{i,c}).$$

Finally, biological effect can be expressed schematically as:

$$P_{D,i} = f(\{k_{i,c}\}, \theta_i),$$

where $\theta_i$ represents microdosimetric and biological context, including reaction-site proximity, subcellular localization, cellular and nuclear geometry, specific-energy distributions, repair capacity, cell-cycle state, and overkill.

This formulation is intentionally modular. Detailed SMK- or PHITS-based calculations can be incorporated into the biological-effect layer, while different imaging, pathological, or agent-specific measurements can be incorporated into the observational layer. The essential distinction is between what is observed, what is inferred, what physically occurs, and how those reactions are translated into biological injury [9,14,15,20–22].

### 4.5. The Aim Is to Infer Distributions, Not Single Values

The primary outputs of the framework should be distributions rather than single summary values. CBE-weighted dose and $D_{\mathrm{isoE}}$ remain useful clinical metrics, but they necessarily suppress heterogeneity that may influence treatment success or failure.

The model should therefore estimate spatial distributions of expected reaction burden, realized reaction counts, cell-death probability, and residual surviving-cell clusters, together with uncertainty intervals. Such outputs can identify regions at risk of reaction insufficiency despite apparently adequate average uptake or dose.

This distributional approach is also compatible with tumor-control-probability-like modeling. Tumor control depends not on the response of the average cell, but on whether clonogenic populations survive and how those populations are spatially organized [9,14,22,26].

**4.6. Tumor and Normal Tissue Must Be Treated in Parallel**

The same hierarchical structure should be applied to both tumor and normal tissues. Tumor modeling estimates the probability of eliminating clonogenic malignant cells, whereas normal-tissue modeling estimates the probability that capture reactions produce clinically meaningful injury in radiosensitive structures.

BNCT selectivity is not guaranteed solely by a higher average boron concentration in tumor. It emerges from the joint distributions of boron uptake, compartmental localization, tissue architecture, neutron fluence, cellular radiosensitivity, and microdosimetric quality across tumor and normal-tissue compartments. A credible effect-prediction framework must therefore model efficacy and toxicity as parallel consequences of the same capture-reaction process [3,25,26].

Normal-tissue inference presents an additional limitation. Because imaging signals such as BPA-PET uptake are often close to background levels in normal tissues, cell-level heterogeneity cannot generally be inferred from imaging alone. Normal-tissue models will therefore depend more heavily on empirically derived priors informed by agent pharmacokinetics, tissue-specific boron measurements, histological architecture, and experimental microdistribution data [20,21,25,27–31].

### 4.7. Position of the Present Proposal

The proposed framework is an inferential bridge rather than a competing dose metric. It connects observable clinical and experimental data to latent cellular variables, reaction-count distributions, and established microdosimetric effect models.

$D_{\mathrm{isoE}}$, SMK-based modeling, PHITS simulation, PET-based heterogeneity correction, MRI-derived structural constraints, and pathology-based morphometry can all be incorporated within this structure. The contribution of the framework is to clarify how these components relate across scales and to propagate uncertainty from macroscopic observations to cellular reaction statistics and biological outcomes.

## 5. Minimal Strategy for Implementation and Validation

### 5.1. Required Inputs

The proposed framework is a blueprint rather than a completed treatment-planning system. Its implementation requires four classes of input that serve as patient-specific observations, region-specific constraints, or empirically informed priors.

The first class comprises imaging-derived inputs, including voxel-level boron-related uptake, spatial heterogeneity, tumor and normal-tissue volumes, and structural or microenvironmental surrogates. In BPA-based BNCT, PET currently provides the most direct clinically accessible information on boron-related uptake, whereas MRI may contribute structural constraints such as cell-size or extracellular-volume surrogates rather than direct boron-localization measurements [20–24].

The second class comprises pathological and histological inputs. Cell density, nuclear size, nuclear-to-cell ratio, stromal fraction, necrosis, and viable tumor fraction define the geometry within which capture reactions occur. Digital pathology may be particularly valuable because it can translate sampled tissue architecture into quantitative priors for the latent cellular layer [10,11,15,25].

The third class comprises boron-agent characteristics, including cellular uptake, tissue selectivity, pharmacokinetics, intracellular retention, cytotoxicity, and compartmental distribution. These data determine how cellular boron burden and intracellular versus extracellular localization should be parameterized. Agent-characterization studies should therefore be regarded as direct sources of mechanistic prior information rather than as separate preclinical exercises [25,27–33].

The fourth class comprises physical and treatment-delivery inputs, including neutron fluence and energy spectrum, irradiation geometry, boron-compound administration, blood boron concentration, treatment timing, and region-specific normal-tissue constraints. These physical quantities remain indispensable; the present framework adds biological, histological, and agent-specific information without displacing conventional dosimetry.

Together, these inputs define the inverse problem. Imaging constrains voxel-level uptake and structure, pathology constrains tissue architecture, agent studies constrain boron microdistribution, and physical modeling constrains the neutron field. Their integration enables inference of latent cellular states and expected $^{10}$B-capture reaction burdens.

### 5.2. Desired Outputs

The primary outputs should extend beyond a single weighted-dose or equivalent-effect value. At minimum, the model should estimate spatial distributions of:

- expected $^{10}$B-capture reaction burden;
- realized reaction-count probabilities;
- cell-death or survival probability;
- residual clusters of surviving malignant cells;
- normal-tissue injury probability; and
- uncertainty associated with each estimate.

These outputs should be defined at least at the voxel or tissue-region level. They may identify regions in which reaction insufficiency persists despite apparently adequate average uptake, absorbed dose, or $D_{\mathrm{isoE}}$. Such regions may reflect voxel-scale cold spots or unresolved cellular heterogeneity within voxels that appear acceptable under conventional metrics.

For tumor tissue, the clinically relevant output is a probabilistic map of residual malignant-cell risk rather than a revised dose map alone. For normal tissue, the corresponding output is the spatial probability of injury arising from the combination of boron distribution, neutron fluence, tissue architecture, and microdosimetric quality.

These distributions may ultimately provide a mechanistic basis for tumor control probability (TCP)- and normal tissue complication probability (NTCP)-like models [26]. Rather than treating control or toxicity as empirical functions of a single aggregate quantity, such models could derive them from spatially heterogeneous reaction burdens and tissue-specific biological responses.

### 5.3. A Four-Stage Validation Strategy

A hierarchical framework of this kind will be persuasive only if its validation is staged and explicit.

The first stage is in vitro validation. Cell systems with characterized boron uptake and compartmental distribution can be used to compare measured surviving fractions with predictions derived from reaction-count distributions and downstream microdosimetric modeling. The initial aim is to test the quantitative relationship among reaction burden, localization, microdosimetric quality, and biological outcome under controlled conditions [7,9,27–31].

The second stage is pathological validation. Inferred distributions of cell density, nuclear size, nuclear-to-cell ratio, stromal content, necrosis, and viable tumor fraction should be compared with measurements from histology and digital pathology. The objective is not exact reconstruction of every cell, but demonstration that the inferred latent states are consistent with observed tissue architecture [10,11,15,25].

The third stage is imaging-consistency calibration and goodness-of-fit assessment. Inferred cellular distributions should remain compatible with the observational constraints provided by PET, MRI, and other available imaging modalities. Systematic disagreement with voxel-level uptake, tumor-to-normal ratio, or structural measurements would indicate that priors, likelihoods, or region-specific parameters require revision [20–24]. This stage constitutes model checking rather than direct validation of unobserved cellular variables.

The fourth stage is clinical-outcome validation. The framework should ultimately be evaluated against local control, recurrence patterns, and normal-tissue toxicity. Its clinical value will depend on whether reaction-count-based hierarchical inference predicts or explains these outcomes better than weighted-dose, CBE-based, or equivalent-effect metrics alone [22,26]. Particular attention should be given to whether inferred reaction-deficient regions correspond to subsequent recurrence and whether microscopic normal-tissue reaction burden improves toxicity prediction.

These stages are cumulative. They progress from controlled reaction-to-effect testing, through validation of latent tissue structure and observational consistency, to demonstration of incremental clinical value.

### 5.4. What the Model Must Outperform

A new framework is justified only if it resolves clinically relevant distinctions that established metrics do not.

First, it should improve the prediction or interpretation of local control and recurrence. Its specific value would be to identify tumors in which mean boron uptake, mean dose, or $D_{\mathrm{isoE}}$ appears adequate, yet low-reaction or reaction-deficient clonogenic populations plausibly remain. Such populations may arise from voxel-scale cold spots, stromal sequestration, viable cells adjacent to necrosis, or unfavorable compartmental localization.

Second, it should improve the prediction or interpretation of normal-tissue toxicity. Microscopic heterogeneity should not be modeled only in tumors. Normal tissues also contain heterogeneous cell populations, geometries, boron distributions, and radiosensitivities. The framework should therefore demonstrate whether parallel modeling of tumor and normal tissue provides a more credible estimate of the therapeutic window than tumor-centered analysis alone.

Third, it should improve decisional resolution. Treatment scenarios that appear equivalent under average-based metrics may differ in reaction-count distributions, uncertainty, residual-risk clusters, or predicted normal-tissue injury. A useful model should distinguish such scenarios and support decisions concerning patient selection, boron-agent development, treatment timing, irradiation planning, or combination therapy.

The framework need not outperform existing metrics in every setting. It must, however, demonstrate added value in cases where spatial or cellular heterogeneity is clinically consequential. If average-based simplification removes structure that determines treatment success or failure, representing that structure is not a theoretical indulgence but a methodological obligation.

### 5.5. Practical Significance

Initial implementations need not reconstruct every target cell individually. The framework can first operate at the voxel or regional level using empirically constrained distributions of boron burden, cellular geometry, and tissue composition. Patient-specific imaging, digital pathology, agent-specific localization data, and refined SMK- or PHITS-based calculations can then be incorporated progressively.

The practical objective is a modular pipeline linking observable data to latent cellular states, reaction-count distributions, microdosimetric quality, and clinical outcomes. Its central product is not a single averaged dose, but a spatially resolved and uncertainty-aware estimate of biological effect.

## 6. Conclusion

BNCT is founded on cell-scale $^{10}B$-capture reactions but is commonly interpreted through macroscopic dose-like quantities. Absorbed dose, CBE-weighted dose, photon-isoeffective dose, and SMK-based modeling remain indispensable, but their biological meaning becomes clearer when the microscopic reaction structure they summarize is made explicit.

We propose cell-associated $^{10}B$-capture reaction burden as an upstream, compartment-resolved variable. Its expected value is determined by boron burden, localization, neutron fluence, and tissue architecture, whereas realized reaction counts remain stochastic and heterogeneous across tumor and normal tissues. Biological effect then emerges from coupling capture-reaction frequency to microdosimetric quality, including reaction-site proximity, specific-energy distribution, cellular geometry, repair capacity, and overkill.

PET, MRI, pathology, boron-agent characterization, and neutron-field modeling provide complementary observational constraints on the latent cellular variables required for this inference. The resulting outputs should therefore be spatial distributions of reaction burden, cell-death probability, residual tumor risk, normal-tissue injury, and uncertainty, rather than a single averaged value.

This framework is not a competing dose metric, but an inferential bridge between established BNCT dosimetry and the cellular logic on which BNCT is founded. Its staged implementation and validation may support more biologically credible and spatially explicit effect prediction.

**Acknowledgement**: The authors are grateful to all members of the Department of Molecular Physiology, Kagawa University Faculty of Medicine, for their helpful discussions and continued support.

**Funding**: none.

**Declaration of generative AI and AI-assisted technologies in the manuscript preparation process**

During the preparation of this work, the authors used ChatGPT (OpenAI) to support manuscript organization, language refinement, and editing. After using this tool, the authors critically reviewed and edited the content as needed and take full responsibility for the content of the published article.

**Conflicts of interest**: none.

**Figure 1**

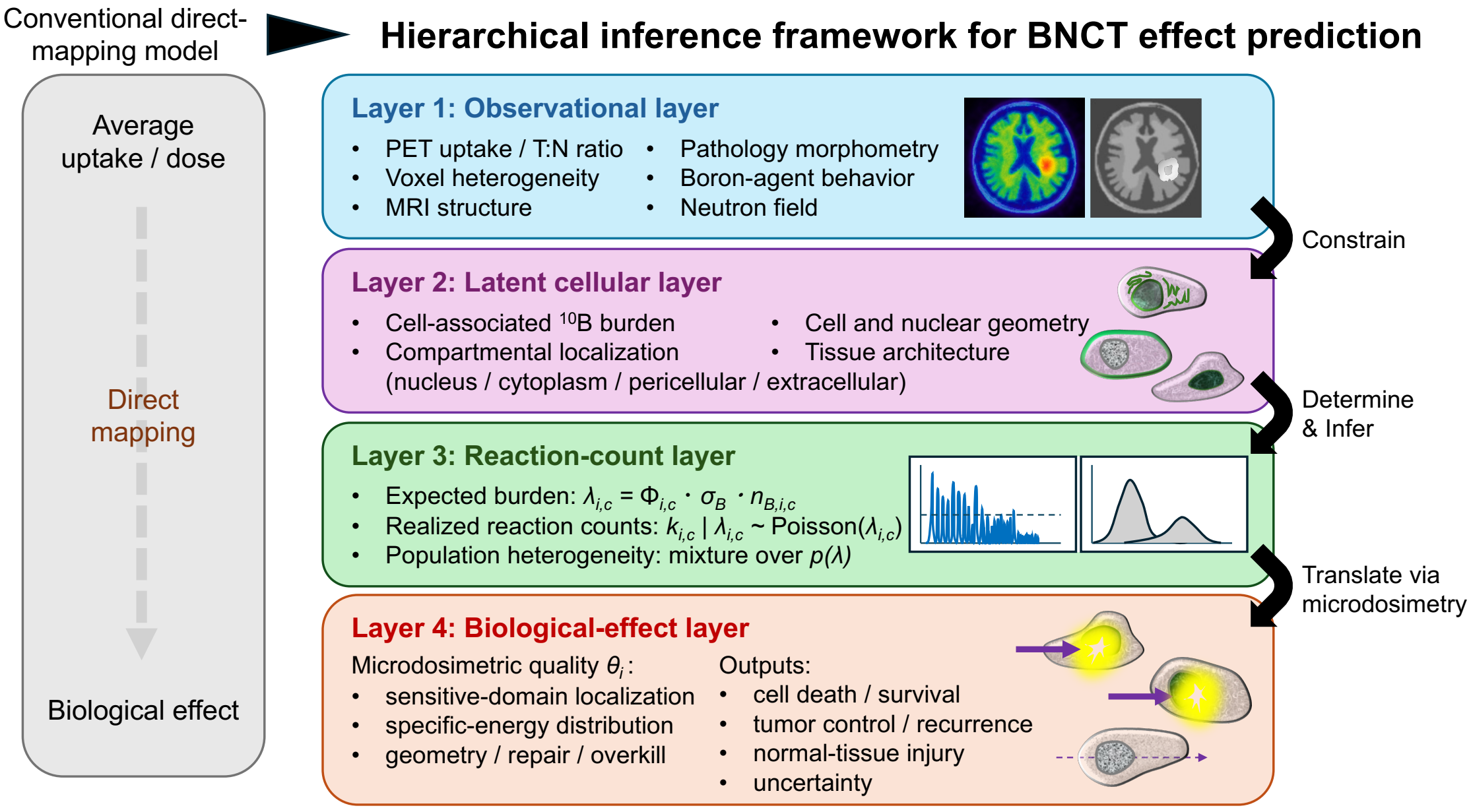

**Primary outputs:** spatial distributions of λ, *k*, cell-death probability, and residual-risk clusters **NOT** a single averaged value.

**Figure 2**

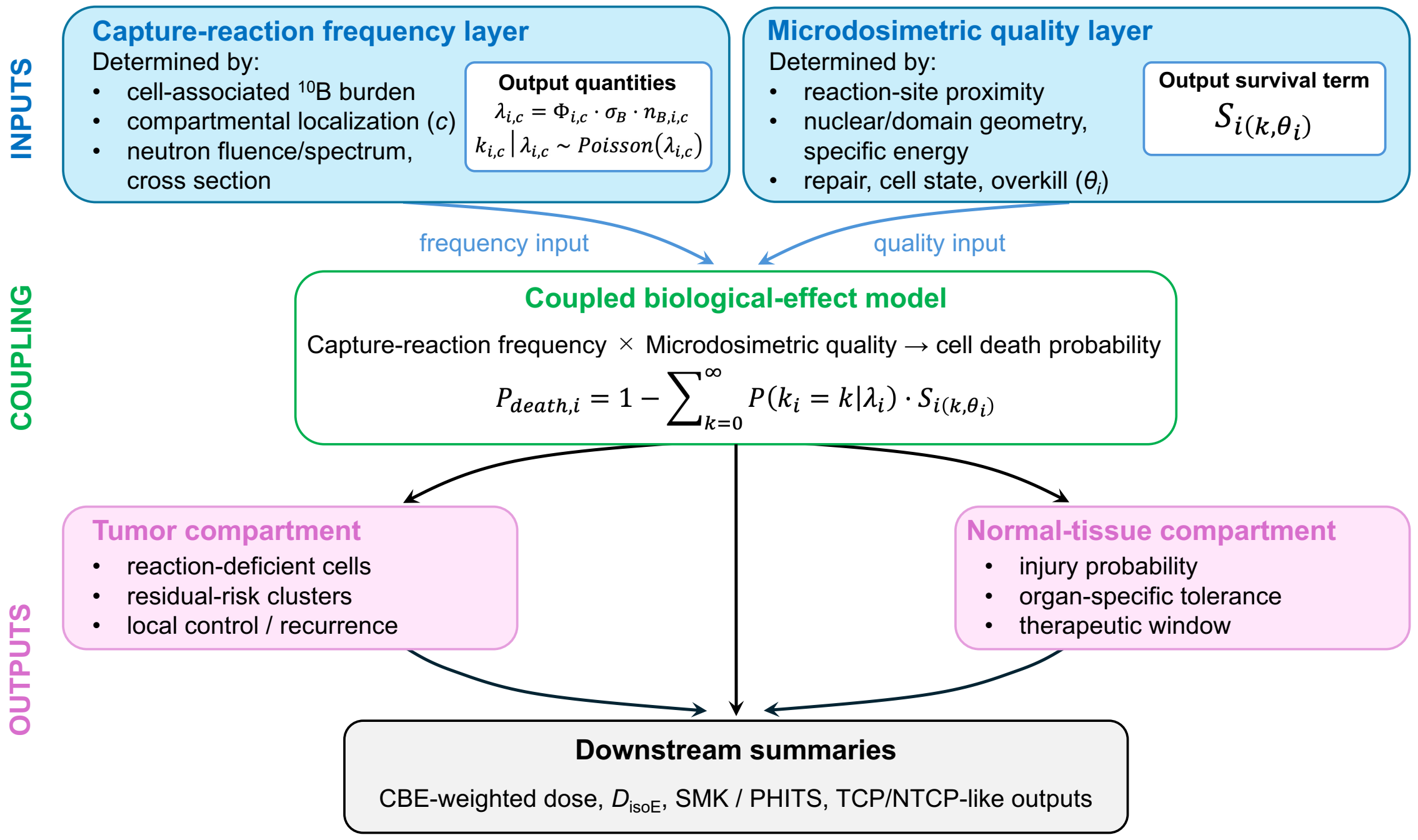
INPUTS
Capture-reaction frequency layer
Determined by:
• cell-associated $^{10}B$ burden
• compartmental localization ($c$)
• neutron fluence/spectrum, cross section
Output quantities
$\lambda_{i,c} = \Phi_{i,c} \cdot \sigma_B \cdot n_{B,i,c}$
$k_{i,c} \mid \lambda_{i,c} \sim Poisson(\lambda_{i,c})$
Microdosimetric quality layer
Determined by:
• reaction-site proximity
• nuclear/domain geometry, specific energy
• repair, cell state, overkill ($\theta_i$)
Output survival term
$S_{i(k,\theta_i)}$
frequency input
quality input
COUPLING
Coupled biological-effect model
Capture-reaction frequency × Microdosimetric quality → cell death probability
$P_{death,i} = 1 - \sum_{k=0}^{\infty} P(k_i = k|\lambda_i) \cdot S_{i(k,\theta_i)}$
OUTPUTS
Tumor compartment
• reaction-deficient cells
• residual-risk clusters
• local control / recurrence
Normal-tissue compartment
• injury probability
• organ-specific tolerance
• therapeutic window
Downstream summaries
CBE-weighted dose, $D_{isoE}$, SMK / PHITS, TCP/NTCP-like outputs

**Figure Legends**:

**Figure 1. Hierarchical inference framework for BNCT effect prediction.**

Clinically and experimentally accessible observations, including PET-derived boron-related uptake, MRI-derived structural information, pathology, boron-agent characterization, and neutron-field data, constrain latent cellular variables such as cell-associated boron burden, compartmental localization, cellular geometry, and tissue architecture. These latent variables determine the expected $^{10}$B-capture reaction burden and the probability distribution of realized reaction counts. Microdosimetric modeling then translates the resulting reaction-count distributions into biological outcomes, including cell death, tumor control, and normal-tissue injury. The framework therefore links macroscopic observations to biological effect through latent cellular states and cell-associated reaction statistics rather than by direct mapping from averaged uptake or dose.

**Figure 2. Coupling of capture-reaction frequency and microdosimetric quality in BNCT effect prediction.**

Cell-associated boron burden, compartmental localization, neutron fluence, and tissue architecture determine the expected $^{10}$B-capture reaction burden, denoted $\lambda_i$ or $\lambda_{i,c}$. The realized reaction count, $k_i$ or $k_{i,c}$, is modeled conditionally on this expected burden, for example using a Poisson distribution. Biological effect depends not only on reaction frequency but also on microdosimetric quality, including reaction-site proximity to radiosensitive domains, microscopic energy deposition, cellular geometry, and biological context. Their coupling determines cell-death probability in tumor and normal-tissue compartments. CBE-weighted dose, $D_{\mathrm{isoE}}$, PHITS-, SMK-, and TCP/NTCP-like quantities are positioned as compatible downstream summaries or modeling outputs rather than as replacements for this upstream structure.

Table 1. Conceptual relationship between existing BNCT effect models and the present framework

| Approach | Primary quantity or concept | What it captures | Remaining limitation | Role in the present framework |
|---|---|---|---|---|
| **Absorbed dose in Gy** | Macroscopic absorbed energy per unit mass | Provides a practical physical dose descriptor | Does not preserve cellular boron heterogeneity, subcellular localization, or discrete capture-reaction statistics | Useful macroscopic summary, but insufficient as a mechanistic descriptor |
| **CBE-weighted dose** | Empirically weighted biological dose | Incorporates compound- and tissue-dependent biological effectiveness | Often relies on fixed or context-dependent weighting factors; limited representation of microscopic heterogeneity | Retained as a practical clinical metric, but interpreted as a downstream summary |
| **Photon-isoeffective dose ($D_{isoE}$)** | Photon dose producing equivalent biological effect | Provides a theoretically grounded equivalent-effect metric | Requires assumptions or inputs regarding microscopic energy deposition and biological response | Positioned as a downstream effect-equivalence output |
| **SMK-based modeling** | Specific-energy distributions and stochastic biological response | Links microdosimetric energy deposition to survival and biological effect | Requires cellular-scale parameters that are not directly observed clinically | Provides the biological-effect layer downstream of inferred reaction statistics |
| **PHITS / cellular-scale simulation** | Particle transport and microscopic dose distribution | Computes detailed energy deposition under defined geometries and boron distributions | Forward simulation; microscopic input parameters must be specified or inferred | Supplies mechanistic priors and downstream microdosimetric modeling |
| **LISMEC** | Interpolated SMK parameters from precomputed cellular-scale simulations | Makes cellular-scale SMK parameter estimation more practical | Still depends on microscopic morphology and boron-distribution parameters | Serves as an implementation-compatible microdosimetric module |
| **PET-based boron imaging** | Voxel-level boron-related uptake and heterogeneity | Provides clinically accessible spatial constraints | Does not directly resolve cell-to-cell or subcellular boron distribution | Forms part of the observational layer |
| **Present framework** | Cell-associated $^{10}$B-capture reaction statistics plus hierarchical inference | Connects imaging/pathology to latent cellular boron distribution, reaction-count statistics, and microdosimetric effect | Conceptual framework requiring staged validation | Provides the inferential bridge between observable data and established effect models |

Box 1. Key terms and notation used in this article

| Term / symbol | Meaning |
|---|---|
| $D_{isoE}$ | Photon-isoeffective dose; the photon dose expected to produce a biological effect equivalent to that produced by BNCT under specified conditions |
| $i$ | Index for a target cell or representative cellular unit |
| $c$ | Spatial compartment, such as nucleus, cytoplasm, pericellular region, or extracellular space |
| $v$ | Imaging voxel or tissue region |
| $n_{B,i,c}$ | Number of $^{10}B$ atoms associated with compartment (c) of target cell (i) |
| $\Phi_{i,c}$ | Time-integrated neutron fluence experienced by compartment (c) of target cell (i) |
| $\sigma_B$ | Effective cross section for the $^{10}B(n,\alpha)^7Li$ reaction |
| $\lambda_{i,c}$ | Expected $^{10}B$-capture reaction burden associated with compartment (c) of target cell (i) |
| $\lambda_i$ | Total expected target-cell-associated reaction burden, summed over relevant compartments |
| $k_{i,c}$ | Realized number of $^{10}B$-capture reactionss associated with compartment (c) |
| $Z_v$ | Latent cellular state within voxel or region (v), including boron distribution, tissue architecture, and cellular geometry |
| $\theta_i$ | Biological and microdosimetric context modifying the consequence of reaction events |
| Cell-associated reaction burden | A target-cell-associated estimate of relevant $^{10}B$-capture events; not restricted to reactions occurring inside the cell nucleus |
| Capture reaction | An occurrence of the $^{10}B(n,\alpha)^7Li$ nuclear reaction |
| Microdosimetric event | Energy imparted to a microscopic target by a single radiation hit; not necessarily identical on a one-to-one basis with a $^{10}B$-capture reaction |